\documentclass[]{article}
\usepackage{comment}
\usepackage{makeidx}  
\usepackage[T1]{fontenc}
\usepackage[latin1]{inputenc}
\usepackage{epic}
\usepackage{eepic}
\usepackage{alltt}
\usepackage{graphicx}
\usepackage{amssymb,amsmath}

\input{psfig.sty}

\usepackage{fullpage}

\def\squarebox#1{\hbox to #1{\hfill\vbox to #1{\vfill}}}

\newcommand{\tb}{\makebox[0.4cm]{}}
\newcommand{\due}{\makebox[0.8cm]{}}
\newcommand{\tre}{\makebox[1.2cm]{}}

\newcommand{\ignore}[1]{}

\newcommand{\ra}{\rangle}
\newcommand{\la}{\langle}
\newcommand\pulse{\mbox{\slshape pulse}}
\newcommand\cycle{\mbox{\slshape cycle}}
\newcommand\Cycle{\mbox{\slshape Cycle}}
\newcommand\cyclemin{\mbox{$cycle_{\mbox{\scriptsize\slshape min}}$}}
\newcommand\cyclemax{\mbox{$cycle_{\mbox{\scriptsize\slshape max}}$}}

\newcommand\proposers{\mbox{\it proposers}}

\newcommand\latest{\mbox{\it latest\_support}}

\newcommand\recent{\mbox{\it recent\_reset}}
\newcommand\cc{\mbox{\it cycle\_countdown}}

\newcommand\PP{``Propose-Pulse''\ }
\newcommand\SP{``Support-Pulse''\ }

\newcommand\RE{``Reset''\ }

\newcommand\ByzAgreement{\mbox{\sc ss-Byz-Agree\ }}
\newcommand\Pulsealg{\mbox{\sc Ab-Pulse-Synch}}

\newcommand\BYZdur{\mbox{$\Delta_{\mbox{\footnotesize\sc byz}}$}}

\newsavebox{\theorembox}
\newsavebox{\lemmabox}
\newsavebox{\conjecturebox}
\newsavebox{\claimbox}
\newsavebox{\factbox}
\newsavebox{\corollarybox}
\newsavebox{\propositionbox}
\newsavebox{\examplebox}
\savebox{\theorembox}{\bf Theorem} \savebox{\lemmabox}{\bf Lemma}
\savebox{\conjecturebox}{\bf Conjecture} \savebox{\claimbox}{\bf
Claim} \savebox{\factbox}{\bf Fact} \savebox{\corollarybox}{\bf
Corollary} \savebox{\propositionbox}{\bf Proposition}
\savebox{\examplebox}{\bf Example}
\newtheorem{theorem}{\usebox{\theorembox}}
\newtheorem{lemma}{\usebox{\lemmabox}}[section]

\newtheorem{corollary}[lemma]{\usebox{\corollarybox}}

\newtheorem{definition}{{\sc Definition}\rm }[section]

\def\squarebox#1{\hbox to #1{\hfill\vbox to #1{\vfill}}}
\newcommand{\qed}{\hspace*{\fill}
\newcommand{\Real}{\mathbb R}
\vbox{\hrule\hbox{\vrule\squarebox{.667em}\vrule}\hrule}\smallskip}

\newenvironment{proof}{\noindent{\bf Proof:~~}}{\(\qed\)}

\begin{document}


\title{Self-Stabilizing Byzantine Pulse Synchronization\\
{\small (revised version)}\footnote{The original version appeared as
TR-2005-84, The Hebrew university, Aug. 2005.}}

\author{
Ariel Daliot\thanks{Email: adaliot@cs.huji.ac.il} \ \ \ Danny
Dolev\thanks{Part of the work was done while the author visited
Cornell University. This research was supported in part by ISF, NSF,
CCR, and AFSOR. Email:
dolev@cs.huji.ac.il} \\School of Engineering and Computer Science,\\
The Hebrew University of Jerusalem, Israel}



\maketitle


\begin{abstract}

The ``Pulse Synchronization'' problem can be loosely described as
targeting to invoke a recurring distributed event as simultaneously
as possible at the different nodes and with a frequency that is as
regular as possible.  This target becomes surprisingly subtle and
difficult to achieve when facing both transient and permanent
failures. In this paper we present an algorithm for pulse
synchronization that self-stabilizes while at the same time
tolerating a permanent presence of Byzantine faults. The Byzantine
nodes might incessantly try to de-synchronize the correct nodes.
Transient failures might throw the system into an arbitrary state in
which correct nodes have no common notion what-so-ever, such as time
or round numbers, and can thus not infer anything from their own
local states upon the state of other correct nodes. The presented
algorithm grants nodes the ability to infer that eventually all
correct nodes will invoke their pulses within a very short time
interval of each other and will do so regularly.

Pulse synchronization has previously been shown to be a powerful
tool for designing general self-stabilizing Byzantine algorithms and
is hitherto the only method that provides for the general design of
efficient practical protocols in the confluence of these two fault
models. The difficulty, in general, to design any algorithm in this
fault model may be indicated by the remarkably few algorithms
resilient to both fault models. The few published self-stabilizing
Byzantine algorithms are typically complicated and sometimes
converge from an arbitrary initial state only after exponential or
super exponential time.

The presented pulse synchronization algorithm will converge by only
assuming that eventually the communication network delivers messages
within bounded, say $d,$ time units, and the number of Byzantine
nodes, $f,$ obeys the $n > 3f$ inequality, for a network of $n$
nodes. The attained pulse synchronization tightness is $3d$ with a
convergence time of a constant number of pulse cycles (each
containing $O(f)$ communication rounds).
\end{abstract}


\thispagestyle{empty}
\newpage
\setcounter{page}{2}


\section{Introduction}
\label{sec:intro}

\textbf{The difficulty of fault tolerant synchronization}:
Coordination and synchronization are among the most fundamental
elements of a distributed task. Nodes typically infer about the
state of the other correct nodes from their own internal states. In
the classic distributed paradigms some extent of initial synchrony
or consistency is always assumed \cite{MinSynchDDS}.
 Even in the classic asynchronous
network model, although nothing is assumed on the time taken for
message delivery, it is typically assumed that nodes have a
controlled and common initialization phase \cite{LynchBook}. Thus it
is assumed that the global state is at least partially consistent
 so that correct nodes have a common notion as to when the system
last initialized. This greatly facilitates the progression of the
algorithm in ``asynchronous rounds'' in which a node knows that if
it has commenced some specific round $r$ then all other correct
nodes have progressed to at least some lesser round. This leads to a
``state-machine replication''
 approach as a general framework to address
the consistency in a distributed environment (see \cite{FTState90}).
Typically, the asynchronous model does not allow for deterministic
fault tolerance as it might not be possible to distinguish between a
late message and a faulty sender (or a lost message). In the
synchronous network model, nodes may assume bounded time for message
delivery (when the system is stable) in addition to assuming that
nodes have a common initialization phase. These two assumptions
allow nodes to use timing criteria to deduce whether certain actions
should have already taken place. This allows for resilience to
permanent faults and plays a pivotal role in the ability to tolerate
Byzantine nodes. Synchronization enables correct nodes to determine
whether a certain message received at a certain time or with a
certain value at this certain time does not agree with the node's
perception of the global progress of the algorithm. In order for all
correct nodes to view symmetrically whether a node does not behave
according to the protocol, it is required to assume that nodes have
similar perceptions of the progress of the algorithm.

A self-stabilizing algorithm does not assume a common initialization
phase. This is required due to transient failures that might corrupt
the local state of nodes, such as the notion as to how long ago the
system or algorithm was initialized. The combination of
self-stabilization and Byzantine fault tolerance poses a special
challenge. The difficulty stems from the apparent cyclic paradox of
the role of synchronization for containing the faulty nodes combined
with the fact that a self-stabilizing algorithm cannot assume any
sort of synchronization or inference of the global state from the
local state. Observe that assuming a fully synchronous model in
which nodes progress in perfect lock-step does not ease this problem
(cf. \cite{DolWelSSBYZCS04}).

 The problem in
general is to return to a consistent global state from a corrupted
global state.  The problem as stated through pulse synchronization,
is to attain a consistent global state with respect to the pulse
event only. I.e. that a correct node can infer that other correct
nodes will have invoked their pulse within a very small time window
of its own pulse invocation. Interestingly enough, this type of
synchronization is sufficient for eventually attaining a consistent
general global state from any corrupted general global state
\cite{ByzStabilizer}. Self-stabilizing Byzantine pulse
synchronization is a surprisingly subtle and difficult problem. To
elucidate the difficulties in trying to solve this problem it may be
instructive to outline a flaw in an earlier attempt to solve this
problem \cite{DDPBYZ-CS03,NASA-BYZSS}: Non-stabilizing Byzantine
algorithms assume that all correct nodes have symmetric views on the
other correct nodes. E.g. if a node received a message from a
correct node then its assumed all correct nodes did so to. Following
transient failures though, a node might initialize in a spurious
state reflecting some spurious messages from correct nodes. With the
pulse synchronization problem, this spurious state may be enough to
trigger a pulse at the node. In order to synchronize their pulses
nodes need to broadcast that they have invoked a pulse or that they
are about to do so. Correct nodes need to observe such messages
until a certain threshold for invoking a pulse is reached. When
nodes invoke their pulses this threshold will be reached again
subsequent to invoking the pulse, causing a correct node to
immediately invoke a pulse again and again.

To prevent incessant pulse invocations, a straightforward solution
is to have a large enough period subsequent to the pulse invocation
in which a node does not consider received messages towards the
threshold. This is exactly where the complimentary pitfall lies,
since some correct nodes may initialize in a state that causes them
to invoke a pulse based on spurious messages from correct nodes. The
consequent pulse message might then arrive at other correct nodes
that initialize in a period in which they do not consider received
messages. Byzantine nodes can, by sending carefully timed messages,
cause correct nodes to invoke their pulses in perfect anti-synchrony
forever. It is no trivial task to circumvent these difficulties.

It is interesting to observe that Byzantine (non-stabilizing) pulse
synchronization can be trivially derived from Byzantine clock
synchronization. Self-stabilizing (non-Byzantine) pulse
synchronization can be easily achieved by following any node that
invokes a pulse. Self-stabilizing Byzantine pulse synchronization on
the other hand is apparently an extremely tricky task.\\

\noindent\textbf{Pulse Synchronization using Byzantine Agreement}:
In our model we do not assume any existing synchrony besides bounded
message delivery. In \cite{FTSS97} it is proven to be impossible to
combine self-stabilization with even crash faults without the
assumption of bounded message delivery. Thus our protocol only
assumes the minimal synchrony required for overcoming crash faults.

The tightly synchronized pulses are produced by utilizing a
self-stabilizing Byzantine agreement protocol, which we have
developed in \cite{DDSSBA-PODC06}, that does not assume any prior
synchronization among the correct nodes. Intuitively, synchronizing
pulses on top of a classic (non-stabilizing) Byzantine agreement
should supposedly be rather straightforward: Execute distributed
Byzantine agreement on the elapsed time remaining until the next
pulse invocation. This scheme requires the correct nodes to
terminate agreement within a short time of each other, but the major
issue is that, unfortunately, when facing transient failures, the
system may end up in a state in which any common reference to time
or even common anchor in time might be lost. This preempts the use
of classic (non-stabilizing) Byzantine agreement and or reliable
broadcast, as these tools typically assume initialization with a
common reference to time or common reference to a round number.
Thus, a common anchor in time is required to execute agreement which
aims at attaining and maintaining a common anchor in time. Thus,
what is required, is an agreement algorithm that is both
self-stabilizing and Byzantine. We resolve this apparent cyclic
paradox by developing in \cite{DDSSBA-PODC06} a self-stabilizing
Byzantine agreement algorithm, named $\ByzAgreement\!,$ with a
unique technique that is based only on the bound on message
transmission time among correct nodes to ``anchor'' a relative time
reference to each invocation of the agreement algorithm. That
algorithm is, to the best of our knowledge, the first Byzantine
agreement algorithm that is also self-stabilizing.

The system may be in an arbitrary state in which the communication
network may behave arbitrarily and in which there may be an
arbitrary number (up-to $n$) of concurrent Byzantine faulty nodes.
The pulse synchronization algorithm will converge once the
communication network eventually resumes delivering messages within
bounded, say $d,$ time units, and the number of Byzantine nodes,
$f,$ obeys the $n > 3f$ inequality, for a network of $n$ nodes. The
attained pulse synchronization tightness is $3d.$ We denote \Cycle\
the targeted time-interval between pulse invocations. The bound on
the effective length of the cycle attained is within $O(d)$ of the
targeted length of $\Cycle.$ The convergence time is 6 cycles (each
containing $O(f)$ communication rounds, where $f$ is the bound on
the number of concurrent permanent faults).\\

\noindent\textbf{Related work}: Pulse synchronization can be
trivially derived from clock synchronization, but no practical
self-stabilizing Byzantine clock synchronization algorithm that does
not assume the existence of synchronized pulses exists.
In~\cite{DolWelSSBYZCS04} the first clock synchronization algorithms
that are self stabilizing and tolerate Byzantine faults are
presented. One of the algorithms assumes a common global pulse and
converges in expected exponential time, the other that doesn't
assume a pulse, converges in expected super-exponential time.
In \cite{DDPBYZ-CS03} we developed an efficient and practical
self-stabilizing Byzantine clock synchronization algorithm based on
pulse synchronization, though the particular pulse synchronization
procedure presented in \cite{DDPBYZ-CS03} suffered from a
flaw\footnote{The flaw was pointed out by Mahyar Malekpour from NASA
LaRC and Radu Siminiceanu from NIA, see \cite{NASA-BYZSS}.}. The
flaw was in neglecting to consider all possible initial values when
the nodes recovers after a transient faults. The current paper
serves as a replacement for that pulse procedure. The clock
synchronization algorithm in \cite{DDPBYZ-CS03} remains largely
unaffected with only a minor change of the clock precision from $3d$
to $4d.$
In \cite{bio-pulse-synch} a novel biologically inspired pulse
synchronization procedure was developed. It has a fundamentally
different structure than the current solution. The current solution
converges in 6 cycles whereas that solution converges in $O(f)$
cycles and has a higher message complexity. Thus, the current
solution scales better with respect to the network size $n.$
In~\cite{Widder-BootCS03} it is shown how to initialize Byzantine
clock synchronization among correct nodes that boot at different
times. Thus eventually they can also produce synchronized Byzantine
pulses (by using the synchronized clocks). That solution is not
self-stabilizing as nodes are booted and thus do not initialize with
arbitrary values in the memory. It has, on the other hand, a
constant convergence time with respect to the required rounds of
communication, whereas our current solution has a dependency on $f,$
which is due to the self-stabilization requirement.
In \cite{ByzStabilizer} it has been shown how, by assuming
synchronized pulses, almost any Byzantine algorithm can be converted
to its self-stabilizing Byzantine counterpart in an efficient and
practical manner. To the best of our knowledge there is sofar no
alternative method besides pulse synchronization for this. That
paper includes a short review on the few other existing
self-stabilizing Byzantine algorithms.

\section{Model and Problem Definition}
\label{sec:model}

The system is a network of $n$ nodes that communicate by exchanging
messages. The nodes regularly invoke ``pulses'', ideally every
$\Cycle$ real-time units. The invocation of the pulse is preceded by
the sending of a message to all the nodes stating the intention of
invoking a pulse. We assume that the message passing allows for an
authenticated identity of the senders. The communication network
does not guarantee any order on messages among different nodes.
Individual nodes have no access to a central clock and there is no
external pulse system. The hardware clock rate (referred to as the
{\em physical timers}) of correct nodes has a bounded drift, $\rho,$
from real-time rate. Consequent to transient failures there can be
an arbitrary number of concurrent Byzantine faulty nodes, the
turnover rate between faulty and non-faulty behavior of the nodes
can be arbitrary and the communication network may behave
arbitrarily. Eventually the system behaves coherently again but in
an arbitrary state.

\begin{definition} A node is {\bf non-faulty} at times that it complies with the
following:
\begin{enumerate}
\vspace{-0.5em} \item \emph{(Bounded Drift)} Obeys a global constant
$0<\rho<<1$ (typically $\rho \approx 10^{-6}$), such that for every
real-time interval $[u,v]:$\vspace{-2mm}
$$(1-\rho)(v-u)  \le \mbox{ `physical timer'}(v) -
\mbox{ `physical timer'}(u) \le (1+\rho)(v-u).$$\vspace{-6mm}

\item \emph{(Obedience)} Operates according to the instructed protocol.

\item \emph{(Bounded Processing Time)} Processes any message of the instructed
protocol within $\pi$ real-time units of arrival time.

\end{enumerate}
\end{definition}

A node is considered {\bf faulty} if it violates any of the above
conditions. We allow for Byzantine behavior of the faulty nodes. A
faulty node may recover from its faulty behavior once it resumes
obeying the conditions of a non-faulty node. In order to keep the
definitions consistent, the ``correction'' is not immediate but
rather takes a certain amount of time during which the non-faulty
node is still not counted as a correct node, although it supposedly
behaves ``correctly''.\footnote{For example, a node may recover with
arbitrary variables, which may violate the validity condition if
considered correct immediately.} We later specify the time-length of
continuous non-faulty behavior required of a recovering node to be
considered \textbf{correct}.

\begin{definition}The communication network is {\bf non-faulty} at periods that it complies with the
following:
\begin{enumerate}
\item Any message
arrives at its destination node within $\delta$ real-time units;

\item The sender's identity and content of any message being received is not
tampered.
\end{enumerate}
\label{def:net_nonfaulty}

\end{definition}

Thus, our communication network model is a ``bounded-delay''
communication network. We do not assume the existence of a broadcast
medium. We assume that the network cannot store old messages for
arbitrary long time or lose any more messages, once it becomes
non-faulty.\footnote{It is enough to assume that messages among
non-faulty nodes are delivered within the specified time bounds.} \\

\newpage
\noindent{\bf Basic definitions and notations: }\\

We use the following notations though nodes do not need to maintain
all of them as variables. To distinguish between a real-time value
and a node's local-time reading we use $t$ for the former and $\tau$
for the latter.
\begin{itemize}
\vspace{-0.5em}

\item $d\equiv \delta + \pi.$ Thus, when the communication network is non-faulty,
$d$ is the upper bound on the elapsed real-time from the sending of
a message by a non-faulty node until it is received and processed by
every non-faulty node.

\item A \textbf{$\pulse$} is an internal event targeted to happen in
``tight''\footnote{We consider $c\cdot d, $ for some small constant
$c,$ as tight.} synchrony at all correct nodes. A \textbf{$\Cycle$}
is the ``ideal'' time interval length between two successive pulses
that a node invokes, as given by the user. The actual \cycle\
length, denoted in regular caption, has upper and lower bounds as a
result of faulty nodes and the physical clock skew.  (Our protocol
requires that $\Cycle >(10f+16)\cdot d.$)

\item $\sigma$ represents the upper bound on the real-time window within which
all correct nodes invoke a pulse ({\em tightness of pulse
synchronization}). We assume that $\Cycle \gg\sigma.$ (Our solution
achieves $\sigma=3d.$)

\ignore{-3mm} \item $\phi_i(t) \in \mathbb{R}^+\cup \{\infty\},$
$0\le i\le n,$ denotes, at real-time $t,$ the elapsed real-time
since the last pulse invocation of $p_i.$ It is also denoted as the
``$\phi$ of node $p_i$''. We occasionally omit the reference to the
time in case it is clear out of the context. For a node, $p_j,$ that
has not sent a pulse since initialization of the system, $\phi_j
\equiv \infty.$

\ignore{\item $\psi_i(t_1,t_2)$ is the number of pulses a correct
node $p_i$ invoked during a real-time interval $[t_1,t_2]$ within
which $p_i$ was continuously correct.}

\item $\cyclemin$ and
$\cyclemax$ are values that define the bounds on the actual \cycle\
length during correct behavior. (We achieve $\cyclemin=\Cycle-11d
\le \cycle \le \Cycle+9d=\cyclemax\;.$)

\ignore{
\ignore{-3mm} \item $\mbox{\slshape message\_decay}\/$
represents the maximal real-time a non-faulty node will keep a
message or a reference to it, before deleting it\footnote{The exact
elapsed time until deleting a messages is specified in the {\sc
prune} procedure in Fig.~\ref{alg:prune}.}.
}

\item \BYZdur\ represents the maximal
real-time required to complete the specific self-stabilizing
Byzantine agreement protocol used. (Using \ByzAgreement in
\cite{DDSSBA-PODC06} it becomes $7(2f+3)d.$)

\end{itemize}

Note that the protocol parameters $n,$ $f$ and $\Cycle$ (as well as
the system characteristics $d$ and $\rho$) are fixed constants and
thus considered part of the incorruptible correct code.\footnote{A
system cannot self-stabilize if the entire code space can be
perturbed, see \cite{CodeStabilizationSSS06}.} Thus we assume that
non-faulty nodes do not hold arbitrary values of these constants. It
is required that $\Cycle$ is chosen s.t. $\cyclemin$ is large enough
to allow our protocol to terminate in between pulses.

A recovering node should be considered correct only once it has been
continuously non-faulty for enough time to enable it to have deleted
old or spurious messages and to have exchanged information with the
other nodes through at least a \cycle.

\begin{definition} The communication network is {\bf correct}
following $\Delta_{net}$ real-time of continuous non-faulty
behavior.\footnote{We will use $\Delta_{net}\ge d.$}
\end{definition}

\begin{definition} A node is {\bf correct} following $\Delta_{node}$  real-time of continuous
non-faulty behavior during a period that the communication network
is correct.\footnote{We will use $\Delta_{node}\ge
\Cycle+\cyclemax.$}
\end{definition}

\begin{definition}\label{def:system-coherence} \emph{(System Coherence)} The system is said to
be {\bf coherent} at times that it complies with the following:

\begin{enumerate}

\item \emph{(Quorum)} There are at least $n-f$ correct nodes,\footnote{The results can be
replaced by $2f+1$ or by $\lceil\frac{n+t}{2}\rceil$ correct nodes.
But for $n>3f+1$ these changes will require some modifications to
the structure of the protocol.} where $f$ is the upper bound on the
number of potentially non-correct nodes, at steady state.

\item \emph{(Network Correctness)} The communication network is correct.
\end{enumerate}
\end{definition}

Hence, if the system is not coherent then there can be an arbitrary
number of concurrent faulty nodes; the turnover rate between the
faulty and non-faulty nodes can be arbitrarily large and the
communication network may deliver messages with unbounded delays, if
at all. The system is considered coherent, once the communication
network and a sufficient fraction of the nodes have been non-faulty
for a sufficiently long time period for the pre-conditions for
convergence of the protocol to hold. The assumption in this paper,
as underlies any other self-stabilizing algorithm, is that the
system eventually becomes coherent.\\

\section{Self-stabilizing Byzantine
Pulse-Synchronization} \label{sec:pulsesynch} \vspace{-2mm} We now
seek to give an accurate and formal definition of the notion of
pulse synchronization. The definitions start by defining a subset of
the system states, called \emph{pulse\_states}, that are determined
only by the elapsed real-time since each individual node invoked a
pulse (the $\phi$'s). Nodes that have ``tight'' or ``close''
$\phi$'s will be called a \emph{synchronized} set of nodes. To
complete the definition of synchrony there is a need to address the
recurring brief time periods in which a node in a synchronized set
of nodes has just invoked a pulse while others are about to invoke
one. This is addressed by considering nodes whose $\phi$'s are
almost a \Cycle\ apart.

If all correct nodes in the system comprise a synchronized set of
nodes then we say that the pulse\_state is a
\emph{synchronized\_pulse\_state of the system}. The goal of the
algorithm is hence to reach a synchronized\_pulse\_state of the
system and to stay in such a state.

\begin{itemize}
\item The {\bf pulse\_state} of the system at real-time $t$ is
given by: $$pulse\_state(t) \equiv (\phi_0(t), \ldots,
\phi_{n-1}(t))\;.$$

\item Let $G$ be the set of all possible pulse\_states of a system
$S.$

\item A set of nodes, $\bar N,$ is called {\bf
synchronized} at real-time $t$ if\\
\vspace{+1mm}$\forall p_i, p_j \in\bar N,$  $\phi_i(t),\phi_j(t) \le
\cyclemax,$ and  one of the following is true:
\begin{enumerate}
\item

 $ |\phi_i(t) - \phi_j(t)| \le \sigma, \mbox{\ \ \ or }$

\item

$ \cyclemin - \sigma \le |\phi_i(t) - \phi_j(t)| \le \cyclemax$ and
$ |\phi_i(t-\sigma) - \phi_j(t-\sigma)| \le \sigma.$
\end{enumerate}

\item $s\in G$ is a {\bf synchronized\_pulse\_state}
\emph{of the system} at real-time $t$ if the set of correct nodes is
synchronized at real-time  $t.$
\end{itemize}

\begin{definition}\label{def:pulse-synch}{\bf The Self-Stabilizing Pulse Synchronization
Problem}\vspace{+2mm}\\

\vspace{-1mm}

\noindent{\bf Convergence:} Starting from an arbitrary system state,
 the system reaches a synchronized\_pulse\_state after a finite
time.\\

\noindent{\bf Closure:} If $s$ is a synchronized\_pulse\_state of
the system at real-time $t_0$ then $\forall\,$real-time $t,t\ge
t_0,$
\begin{enumerate}
\vspace{-1mm} \item pulse\_state(t) is a synchronized\_pulse\_state,

\ignore{\item $(t-t_0\le \cycle_{\mbox{\scriptsize\slshape  min}}
\Rightarrow \psi_i(t,t_0)\le 1) \bigwedge (t-t_0\ge
\cycle_{\mbox{\scriptsize\slshape  max}} \Rightarrow
\psi_i(t,t_0)\ge 1),$\\ for every correct node $i.$}

\item In the real-time interval [$t_0,\;t$] every correct node will invoke at most a single pulse if $t-t_0\le\cyclemin$
and will invoke at least a single pulse if $t-t_0\ge\cyclemax.$
\end{enumerate}
\end{definition}

The second Closure condition intends to tightly bound the effective
pulse invocation frequency within a priori bounds. This is in order
to defy any trivial solution that could synchronize the nodes, but
be completely unusable, such as instructing the nodes to invoke a
pulse every $\sigma$ time units. Note that this is a stronger
requirement than the ``linear envelope progression rate'' typically
required by clock synchronization algorithms, in which it is only
required that clock time progress as a linear function of
real-time.

\subsection{The Pulse Synchronization Algorithm}
\vspace{-2mm} The self-stabilizing Byzantine pulse synchronization
algorithm presented is called $\Pulsealg$ (for
\textit{Agreement-based Pulse Synchronization}). A \cycle\ is the
time interval between two successive pulses that a node invokes. The
input value $\Cycle$\/ is the ideal length of the \cycle. The actual
real-time length of a \cycle\ may deviate from the value $\Cycle$\/
in consequence of the clock drifts, uncertain message delays and
behavior of faulty nodes. In the proof of Lemma~\ref{clm:all-pp} the
extent of this deviation is explicitly presented.

The environment is one without any granted synchronization among the
correct nodes besides a bound on the message delay. Thus, it is of
no use whether a sending node attaches some time stamp or round
number to its messages in order for the  nodes to have a notion as
to when those messages supposedly were sent. Hence in order for all
correct nodes to symmetrically relate to any message disseminated by
some node, a mechanism for agreeing on which phase of the algorithm
or ``time'' that the message relates to must be implemented. This is
fulfilled by using \ByzAgreement\!\!, a self-stabilizing Byzantine
agreement protocol presented in \cite{DDSSBA-PODC06}. The mode of
operation of this protocol is as follows: A node that wishes to
initiate agreement on a value does so by disseminating an
initialization message to all nodes that will bring them to
(explicitly) invoke the \ByzAgreement protocol. Nodes that did not
invoke the protocol may join in and execute the protocol in case
enough messages from other nodes are received during the protocol.
The protocol requires correct initiating nodes not to disseminate
initialization messages too often. In the context of the current
paper, a \SP message serves as the initialization message.

When the protocol terminates, the protocol \ByzAgreement returns at
each node $q$ a triplet $(p,m,\tau^p_q),$ where $m$ is the agreed
value that $p$ has sent. The value $\tau^p_q$ is an estimate, on the
receiving node $q$'s local clock, as to when node $p$ have sent its
value $m.$ We also denote it as the ``recording time'' of $(p,m).$
Thus, a node $q$'s decision value is $\la p,m,\tau^p_q\ra$ if the
nodes agreed on $(p,m).$ If the sending node $p$ is faulty then some
correct nodes may agree on $(p,\perp),$ where $\perp$ denotes a
non-value, and others may not invoke the protocol at all. The
function $rt(\tau_q)$ represents the real-time when the local clock
of $q$ reads $\tau_q.$ The $\Pulsealg$ algorithm uses the
\ByzAgreement protocol for a single message only (\SP message) and
not for every message communicated. Thus the agreement is on whether
a certain node sent a \SP message and when, and not on any actual
value sent. Correct nodes do not send this message more than once in
a \cycle.

The \ByzAgreement protocol satisfies the following typical Byzantine
agreement properties:\\


\noindent{\bf Agreement:} If the protocol returns a value
($\neq\perp$) at a correct nodes, it returns the same value at all
correct nodes;

\noindent{\bf Validity:} If all correct nodes are triggered to
invoke the protocol \ByzAgreement by a value sent by a correct node
$p,$ then all correct nodes return that value;

\noindent{\bf Termination:} The protocol terminates in a finite
time;

\vspace{+2mm}It also satisfies some specific timeliness properties
that are listed in Section~\ref{sec:proof}.\\

The heuristics behind \Pulsealg\  protocol are as following:

\begin{itemize}

\vspace{-2mm}\item Once the node approaches its end of \Cycle, as
measured on its physical timer, it sends a \PP message stating so to
all nodes.

\vspace{-2mm}\item When ($n-f$) distinct \PP messages are collected,
the node sends a \SP message that states so to all nodes. This
serves as the initialization message for invoking agreement.

\vspace{-2mm}\item Upon receiving such a message a receiving node
invokes self-stabilizing Byzantine agreement (\cite{DDSSBA-PODC06})
on the fact that it received such a message from the specific node.
We require that \Cycle\ be long enough to allow the agreement
instances to terminate.

\vspace{-2mm}\item If all correct nodes invoked agreement on the
same message  within a short time window then they will all agree
that the sender indeed sent this \SP message and all will have
proximate estimates as of when that node could have sent this
message.

\vspace{-2mm}\item The time estimate is then used to reset the
countdown timer for the next pulse invocation and a consequent
``reset'' messages to be sent. Each new agreement termination causes
a renewed reset.

\vspace{-2mm}\item Upon arrival of a reset message the sending  node
is taken off the list of nodes that have ended their \Cycle\  (as
indicated by the earlier arrival of a \PP  message for that  node).

\vspace{-2mm}\item Thus, some short time after all correct nodes
have done at least one reset of  their cycle countdown timer, no new
agreement can be initiated by  any node (faulty or correct).

\vspace{-2mm}\item Thus, there is one agreement termination that
marks a small  time-window within which all correct nodes do a last
reset of the cycle countdown  timer. Thus, essentially, all correct
nodes have synchronized the invocation of  their next pulse.
\end{itemize}

The algorithm is executed in an ``event-driven'' manner. Thus, each
node checks the conditions and executes the steps (blocks) upon an
event of receiving a message or a timer event. To simplify the
presentation it is assumed in the algorithm that when a correct node
sends a message it receives its own message through the
communication network, like any other correct node.

The algorithm assumes a timer that measures interval of time of size
$\Cycle.$ The algorithm uses several sets of messages or references
that are reset throughout the algorithm, and every message that have
arrived more than $\Cycle+2d$ ago is erased.

\begin{figure}[ht!]
\center \fbox{\begin{minipage}{5.7in} \footnotesize
\setlength{\baselineskip}{2.9mm}

\noindent Algorithm $\Pulsealg$\! \emph{(n, f, \Cycle)}\hfill\textit{/* continuously executed at node $q$ */}\\
\\
A1.\ {\bf if} ($cycle\_countdown=0$) {\bf then}
\mbox{\ }\hfill\textit{/* assumes a background process}\\\mbox{\ }\hfill\textit{ that continuously reduces  \cc\ */} \\
A2.\ \tb     $cycle\_countdown:=\Cycle$; \\
A3.\ \tb     send \PP message to all; \hfill\textit{/* endogenous message */} \\
\\\vspace{1.5mm}\\\noindent
B1.\ {\bf if} received \PP message from a sender $p$ and $p\not\in\recent_q$ {\bf then}\\
B2.\ \tb add $p$ to $\proposers_q$;
\\\\\vspace{1.5mm}\\\noindent
C1.\ {\bf if} $q\in \proposers_q\;\&\;\|\proposers_q\| \ge n-f$  and\\
C2.\ \tb did not send a \SP in the last $\Cycle-8d$ {\bf then}\\
C3.\ \due send ``Support-Pulse($\proposers_q$)'' to all;\hfill\textit{/* support the forthcoming pulse */}\\
\\\vspace{1.5mm}\\\noindent
D1.\ {\bf if} received ``Support-Pulse($\proposers_p$)'' message from a sender $p$ and in the last $\Cycle-11d$\\
D2.\ \tb did not invoke \ByzAgreement($p,$``support'') or decide on $\la p,$``$support$'',$\_\ra$  and\\
D3.\ \tb within $d$ of its reception $\|(\proposers_q\cup\recent_q)\cap\proposers_p\|\ge f+1$ {\bf then}\\
D4.\ \due \ByzAgreement\!($p,$ ``$support$'')\hfill\textit{/* invoke agreement on the pulse supporter */};\\
\\\vspace{1.5mm}\\\noindent
E1.\ {\bf if} \textbf{decided} on $\la
p,\;$``$support$'',$\;\tau^p_q\ra$ at some local-time $\tau_q$ {\bf
then}
\hfill\textit{/* on non $\perp$ value */} \\
E2.\ \tb{\bf if}  $\tau^p_q\ge \latest_q$  {\bf then} \\
E3.\ \due $\latest_q:=\tau^p_q$;    \hfill\textit{/*  the latest agreed supporter so far at q */}\\
E4.\ \due {\bf if} not invoked a pulse since local-time $\tau_q-(\BYZdur+6d)$ {\bf then} \hfill\textit{/* pulse separation */}\\
E5.\ \tre    {\bf invoke} the $\pulse\/$ event;\\
E6.\ \due $cycle\_countdown := \Cycle-(\tau_q-\tau^p_q)$;\hfill\textit{/* reset cycle */}\\
E7.\ \due send \RE message to all and remove yourself, $q,$ from $\proposers_q$;\\
\\\vspace{1.5mm}\\\noindent
F1.\ {\bf if} received \RE from a sender $p$  {\bf then}\\
F2.\ \tb move $p$ from  $\proposers_q$ to $\recent_q$;\hfill\textit{/* \recent\/ decay within $2d+\epsilon$ time */}\\
\\\vspace{1.5mm}\\\noindent
\mbox{\ }\ Continuously ongoing cleanup: \\
G1.\ \tb delete an older message if a subsequent one arrives from
the same sender;\\
G2.\ \tb delete any data in $\recent_q$ after $2d+\epsilon$ time units; \\
G3.\ \tb reset \cc\ to be \Cycle\, if $\cc\not\in[0,\Cycle]$; \\
G4.\ \tb reset $\latest_q$ to be $\tau-\Cycle$ if $\latest_q\not\in[\tau-\Cycle,\tau]$; \\
G5.\ \tb delete any other message or data that is older than
$\Cycle+2d$ time units;
\normalsize
\end{minipage} }
\caption{The $\Pulsealg$ Pulse Synchronization Algorithm}
\label{alg:pulse}
\end{figure}

The algorithm assumes the ability of nodes to estimate some time
intervals, like at Line~C2. These estimates can be carried out also
in a self-stabilizing environment, by tagging each event according
to the reading of the local timer.  So even if the initial values
are arbitrary and cause the non-faulty node to behave
inconsistently, by the time it is considered correct the values will
end up resetting  to the right values. Note that the nodes do not
exchange clock values, rather they measure time locally on their own
local timers. It is assumed that a non-faulty node handles the wrap
around of its local timer while estimating the time intervals.

Note that there is no real reason to keep a received message after
it has been processed and its sender been referred to in the
appropriate data structures. Hence, if messages are said to be
deleted after a certain period, the meaning is to the reference of
the message and not the message itself, which can be deleted
subsequent to processing.

For reasons of readability we have omitted the hardware clock skew
$\rho,$ from the constants, equations and proofs. The introduction
of $\rho$ does not change the protocol whatsoever nor any of the
proof arguments. It only adds a small insignificant factor to many
of the bounds.\\

We now seek to explain in further detail the blocks of the
algorithm:\\

Block A: We assume that a background process continuously reduces
the counter $cycle\_countdown,$ intended to make the node count
$\Cycle$\/ time units on its physical timer. On reaching $0,$ the
background process resets the value back to $\Cycle.$ It expresses
its intention to synchronize its forthcoming pulse invocation with
the pulses of the other nodes by sending an endogenous {\bf \PP}
message to all nodes. Note that a reset is also done if
$cycle\_countdown$ holds a value not between $0$ and $\Cycle.$ The
value of $cycle\_countdown$ is also reset once the ``pulse'' is
invoked. Observe that nodes typically send more than one message in
a \cycle, to prevent cases in which the system
may be invoked in a deadlocked state.\\

Block B: The \PP messages are accumulated at each correct node in
its $\proposers$ set. We say that two messages are {\bf distinct} if
they were sent by different nodes.\\

Block C: These messages are accumulated until enough (at least
$n-f$) have been collected. If in addition the node has already
proposed itself then the node will declare this event through the
sending of a {\bf \SP} message, unless it has already sent such a
message not long ago.  The message bears a reference to the nodes in
the $\proposers$ set of the sender. Note that a node that was not
able to send the message because sending one not long ago, may send
it later when the conditions will hold.\\

Block D: Any such \SP message received is then checked for
credibility by verifying that the history it carries has enough (at
least $f+1$) backing-up in the receiver's $\proposers$ set and that
a previous message was not sent recently. It is only then that
agreement is initiated, on a credible pulse supporter. Note that a
correct node would not have supported a pulse (sent a \SP message)
unless it received $n-f$ propose messages and has not sent one
recently. Thus all correct nodes will receive at least $f+1$ propose
messages from correct nodes and will join the agreement initiation
by the pulse
supporter within $d$ real-time units.\\

Block E: The Byzantine agreement protocol decides whether a certain
node issued a \SP message. Each node $q$ decides at some local-time
$\tau_q.$ The agreement protocol also returns an estimate as of
when, on the deciding node's local clock, the message was sent by
the initiating node. This time is denoted $\tau^p_q.$ Correct nodes
end up having bounded differences in the real-time translation of
their $\tau^p$ values, for a specific agreement.

When a node decides on a value it checks whether the $\tau^p$
returned by the agreement protocol is the most recent decided on so
far in the current \cycle. Only then are lines E3-E7 executed. Note
that the same agrement instance may return a $\tau^p,$ which is the
most recent one for a certain correct node but may not be the most
recent at another correct node. This can happen because correct
nodes terminate the \ByzAgreement protocol within $3d$ time units of
each other,\footnote{It is part of the timeliness properties of the
\ByzAgreement protocol, see Section~\ref{sec:proof}.} and their
translation of the realtime of the $\tau^p$ values may differ by
$5d.$ Thus, this introduces a $3d$ time units uncertainty between
the execution of the subsequent lines at correct nodes.

In Line E4-E5 a pulse is invoked if no pulse has recently been
invoked. In Line E6 the node now resets the \cycle\ so that the next
pulse invocation is targeted to  happen at about one $\Cycle$\/
later. In Line E7 a \RE message is sent to all nodes to inform that
a reset of the \cycle\ has been done. The function of this message
is to make every node that resets, be taken out of the $\proposers$
set of all other correct nodes\footnote{Note that a node may send
multiple \RE messages. It is done in order to simplify some of the
claims in the proof.}. To ensure that only one pulse is invoked in
the minimal time span of a \cycle\ a pulse will not be invoked in
Line E4 if
done so recently.\\

Block F: This causes all correct nodes to eventually remove all
other correct nodes from their $\proposers.$ Thus, about $2d$ after
all correct nodes have executed Line E7 at least once, no instance
of \ByzAgreement will be initiated by any correct node and
consequently no more agreements can terminate (beyond the currently
running ones). The last agreement decision of the correct nodes,
done within a short time-window of each other, returns different but
closely bounded $\tau^p$ values at the correct nodes. Consequently
they all reset their $cycle\_countdown$ counters to proximate
values. This yields a quiescent window between the termination of
the last agreement and the next pulse invocation, which will be
invoked
within a small time window of each other.\\

Block G: The scheme outlined above is not sufficient to overcome the
cases in which some nodes initialize with reference to spurious
messages sent by other nodes while such messages were not actually
sent. The difficulty lies in the fact that Byzantine nodes may now
intervene and constantly keep the correct nodes with asymmetric
views on the sets of messages received. To overcome this,
$\Pulsealg$ has a decay process in
which each data that is older than some period is deleted.

Note that the decaying of values is carefully done so that correct
nodes never need to consider messages that arrived more than $\Cycle+2d$ ago.\\

\vspace{-3mm}\subsection{Proof of Correctness}
\label{sec:proof}

The proof of correctness requires very careful argumentation and is
not a straightforward standard proof of the basic properties. The
critical parts in the proof is showing that despite the complete
chaotic initialization of the system the correct nodes are able to
produce some relation among their local clocks and force the faulty
nodes to leave a short interval of time into which no recording time
refers to, followed by an interval during which no correct node
updates its \latest. After such intervals we can argue about the
convergence of the states of the correct nodes, proving that
stability is secured. The nontraditional values of the various
constants bounding \Cycle\ has to do with the balance between
ensuring the ability to converge and limiting the ability of the
Byzantine nodes to disturb the convergence by introducing critically
timed pulse events that may disunite the correct nodes.

The proof shows that when the constants are chosen right, no matter
what the faulty nodes will do and no matter what the initial values
are, there will always be two intervals of inactivity, concurrently
at all correct nodes, after which the correct nodes restore
consistency of their pulses.

The proof uses the following specific properties of the
\ByzAgreement protocol (\cite{DDSSBA-PODC06}):\\

\vspace{-2mm}\noindent{\bf Timeliness-Agreement Properties:}
\vspace{-2mm}\begin{enumerate}
\item (agreement)
For every two correct nodes $q$ and $q^\prime$ that decides $\la
p,m,\tau^p_q \ra$ and $\la p,m,\tau^p_{q^\prime} \ra$ at local times
$\tau_q$ and $\tau_{q^\prime},$ respectively:
\begin{enumerate}
\item
$|rt(\tau_q)-rt(\tau_{q^\prime})|\le 3d,$ and if validity holds,
then $|rt(\tau_q)-rt(\tau_{q^\prime})|\le 2d.$
\item
$|rt(\tau^p_q)-rt(\tau^p_{q^\prime})|\le 5d.$
\item
$rt(\tau^p_q), rt(\tau^p_{q^\prime}) \in [t_1-2d,t_2],$ where
$[t_1,t_2]$ is the interval within which all correct nodes that
actually invoked \ByzAgreement$\!(p,m)$  did so.
\item $rt(\tau^p_q)\le rt(\tau_q)$ and $rt(\tau_q)-rt(\tau^p_q)\le \BYZdur$ for every correct node
$q.$
\end{enumerate}

\item (validity)
If all correct nodes invoked the protocol in an interval
$[t_0,t_0+d],$ as a result of some initialization message containing
$m$ sent by a correct node $p$ that spaced the sending by at least
$6d$ from the completion of the last agreement on its message, then
for every correct node $q,$ the decision time $\tau_q,$ satisfies
$t_0-d \le rt(\tau^p_q)\le rt(\tau_q)\le t_0+3d.$

\item (separation) Let $q$ be any correct node that decided on any
two agreements regarding $p$ at local times $\tau_q$ and
$\bar\tau_q,$ then $t_2+5d<\bar t_1$ and $rt(\tau_q)+5d< \bar
t_1<rt(\bar\tau_q),$ where $t_2$ is the latest real-time at which a
correct node invoked \ByzAgreement in the earlier agreement and
$\bar t_1$ is the earliest real-time that \ByzAgreement was invoked
by a correct node in the later agreement.

\end{enumerate}

The \Pulsealg\ requires the following bounds on the variables:

\begin{itemize}
\item $\Cycle\ge\max[(10f+16)d,\;\BYZdur+14d].$
\item $\Delta_{node}\ge
\Cycle+\cyclemax.$
\item $\Delta_{net}\ge d.$
\end{itemize}

The requirements above, and the definitions of correctness imply
that from an arbitrary state the system becomes coherent within 2
cycles.

Note that in all the theorems and lemmata in this paper, if not
stated differently, it is assumed that the system is coherent, and
the claims hold as long as the system stays coherent.

In the proof, whenever we refer to correct nodes that decide we
consider only decisions on $\neq\perp$ values. When the agreement
returns $\perp$ it is not considered a decision, and in such a case
the agreement at other correct nodes may not return anything or may
end up in decaying all related messages.

\begin{theorem}\label{thm:converge}
(\emph{Convergence}) From an arbitrary (but coherent) state a
synchronized\_pulse\_state is reached within 4 cycles, with
$\sigma=3d.$
\end{theorem}

\begin{proof}{A node that recovers may find
itself with arbitrary input variables and in an arbitrary step in
the protocol. Within a \cycle\ a recovered node will decay all
spurious ``messages'' that may exist in its data structures. Some of
these might have been resulted from incorrect initial variables,
such as when invoking the \ByzAgreement protocol without the
specified pre-conditions. Such effects also die out within a \cycle.

The above argument implies that by the time the node is considered
correct, all messages sent by non-faulty nodes that are reflected in
its data structures were actually sent by them (at the arbitrary
state at which they are). Thus, by the time that the system becomes
coherent the set of correct nodes share the values they hold in the
following sense: if a message sent by a non-faulty node is received
by a correct node, then within $d$ it will be received by all other
correct nodes; and all future messages sent by correct nodes are
based on actual messages that were received.

Once the system is coherent, then there are at least $n-f$ correct
nodes that follow the protocol, and all messages sent among them are
delivered by the communication network and processed by the correct
nodes within $d$ real-time units.

\begin{lemma}\label{clm:pp-msg}
Within $d$ real-time units of the sending of a \PP message by a
correct node $p,$ it appears in  $\proposers_q$ of any correct node
$q.$ Furthermore, it appears in $\proposers_q$ only if $p$ sent a
\PP message within the last $d$ units of time.
\end{lemma}

\begin{proof}
From the coherence of the system, $p$'s message arrives to all
within $d$ real-time. By the Timeliness-Agreement Property (1d) and
the bounds on \Cycle, a node that have recently sent a \RE message
resets its \cc\ to a value that is at least $\Cycle-\BYZdur>14d.$
Thus, the minimum real-time between the receipt of its past \RE and
its current \PP at any correct node is more than $2d$ apart, and
therefore by the time its \PP message arrives it will not appear in
$\recent_q$ at any correct node $q.$ The second part is true because
$p$ can be in $\proposers_q$ without prior sending of a \PP message
only if node $q$ recovered in that state. But by the time node $q$
is considered correct any reference to such a message has already
been decayed.
\end{proof}

\begin{lemma}\label{clm:n-t-in-cycle} In every real-time interval
equal to \Cycle, every correct node  sends either a \PP message or a
\RE message.
\end{lemma}

\begin{proof}
Recall that every correct node's $cycle\_countdown$ timer is
continuously running in the background and would be reset to hold a
value within $\Cycle$ if it initially held an out-of-bound value.
Thus, if the $cycle\_countdown$ is not reset to a new value when a
\RE is invoked, then within $\Cycle$ real-time units the
$cycle\_countdown$ timer will eventually reach $0$ and a \PP message
will consequently be sent. Whenever a $cycle\_countdown$ is reset,
its value is always at most \Cycle.
\end{proof}

\begin{lemma}\label{clm:delete-pp-msg}
Within $d$ real-time units of sending a \RE message by a correct
node $p,$ that node does not appear in  $\proposers_q$ of any
correct node $q.$ Furthermore, a correct node $p$ is deleted from
$\proposers_q$ only if it sent a \RE message.
\end{lemma}

\begin{proof}
The first part follows immediately from executing the protocol in a
coherent state. The only sensitive point arises when a \PP message
that was sent by $p$ prior to the \RE message arrives after the \RE
message. This can happen only if the \PP message was sent within $d$
of the \RE message. But in this case the protocol instructs node $q$
not to add $p$ to $\proposers_q.$  For proving the second part we
need to show that a correct node is not removed from $\proposers_q$
because $q$ decayed it.  By Lemma~\ref{clm:pp-msg} it appears in
$\proposers_q$ only because of sending a \PP message.  By
Lemma~\ref{clm:n-t-in-cycle} it will resend a new message before $q$
decays the previous message, because messages are decayed (Block G)
only after $\Cycle+d.$
\end{proof}

\begin{lemma}\label{clm:support-correct} Every correct node
invokes \ByzAgreement$(p,``support$''$)$ within $d$ real-time units
of the time a correct node $p$ sends a \SP message.
\end{lemma}

\begin{proof}
If a correct node $p$ sent a \SP message in Line~C3, then the
preconditions of Line~D2 hold because the last reception of \SP and
the last invocation of \ByzAgreement$(p,``support$''$)$ that
followed took place at least $\Cycle-8d-d$ ago, proving the first
condition. By the Timeliness-Agreement property~(2) the last
decision took place at least $\Cycle-8d-3d$ ago, proving the other
condition. The condition in Line D3 clearly holds for all correct
nodes. This is because within $d$ real-time units every correct node
in $\proposers_p$ will appear in $\proposers_q$ and every correct
node that was deleted from $\proposers_q$ and is not in $\recent_q$
should have been already deleted from $\proposers_p.$ To prove this
last claim, assume that node $q$ received \RE from a correct node
$v$ at real-time $t.$ By $t+d$ this message should arrive at $p,$
and therefore any \PP message from $p$ that contains $v$ should be
sent before that and should be received before $t+2d,$ thus before
removing $v$ from $\recent_q.$
\end{proof}

\begin{lemma}\label{clm:minus-d}
If a correct node $p$ sends \SP  at real-time $t_0$ then every
correct node $q$ decides $\la p,$``support'',$\tau^p_q\ra$ at some
local-time $\tau_q,$ such that $t_0-d\le rt(\tau^p_q) \le
rt(\tau_q)\le t_0+3d$ and $t_0 \le rt(\tau_q).$
\end{lemma}

\begin{proof}
By Lemma~\ref{clm:support-correct} all correct nodes invoke
\ByzAgreement$(p,``support$''$)$ in the interval $[t_0,t_0+d].$ Thus
the precondition conditions for the Timeliness-Agreement
property~(2) hold. Therefore, each correct node $q$ decides on $\la
p,\_,\tau^p_q\ra$ at some real-time $rt(\tau_q)$ that satisfies
$t_0-d\le rt(\tau^p_q) \le rt(\tau_q)\le t_0+3d.$
\end{proof}

\begin{lemma}\label{clm:decide on-exists}
Let $[t,t+\Cycle]$ be an interval such that for no correct node
$rt(\latest)\in[t,t+\Cycle],$ then by $t+\Cycle+4d$ all correct
nodes decide.
\end{lemma}

\begin{proof}
Assume that all decisions by correct nodes resulted in
$rt(\latest)\le t.$ Thus, since there are no updates to \cc\, the
\cc\ at all correct nodes should expire by $t+\Cycle.$ By
Lemma~\ref{clm:minus-d}, if any correct node would have sent \SP in
that interval, then we are done. Otherwise, by that time all should
have sent a \PP message. Since no node removes old messages for
$\Cycle+2d,$ and more than $\Cycle-8d$ real-time passed, by
$t+\Cycle + d$ at least one correct node will send a \SP message. By
Lemma~\ref{clm:support-correct}, all will invoke \ByzAgreement
within another $d$ real-time units. The Timeliness-Agreement
property~(2) implies that by $t+\Cycle + 4d$ all will decide.
\end{proof}

Note that if a faulty node sends \SP\!\!, some correct node may join
and some may not, and the actual agreement on a value $\neq\perp$
and the time of such an agreement depends on the behavior of the
faulty nodes. We address that later on in the proof. We first prove
a technical lemma.

\begin{lemma}\label{clm:6d}
Let $t^\prime,$ be a time by which all correct nodes decided on some
values since the system became coherent. Let $B^\prime$ and $B$
satisfy $B^\prime\le B,$ and $3d\le B.$ If no correct node decides
on a value that causes updating $\latest$ to a value in an interval
$[t^\prime,t^\prime+B],$ and no correct node updates its \latest\ or
resets its \cc\ during the real-time interval
$[t^\prime+B^\prime,t^\prime+B],$ then for any pair of correct nodes
$|\cc_q(t^{\prime\prime})-\cc_{q^\prime}(t^{\prime\prime})|\le 5d$
for any $t^{\prime\prime},$ $t^\prime+B^\prime\le\
t^{\prime\prime}\le t^\prime+B.$
\end{lemma}

\begin{proof} By assumption, the agreements prior to $t^\prime$
satisfy the Timeliness-Agreement properties. Past
$t^\prime+B^\prime$ and until $t^\prime+B$ no node updates its
\latest. Thus, for all nodes the value of $rt(\latest)$ is bounded
by $rt(\latest)\le t^\prime.$ Let $q$ be the correct node with the
maximal $rt(\latest_q)$ that was set following a decision $\la
p_1,\_,\tau_1^{p_1}\ra$ at timer $\tau_1,$ where
$\latest_q=\tau_1^{p_1}.$ By the Timeliness-Agreement property~(1a),
any correct node $v$ will execute Line E2 following a decision on
$\la p_1,\_,\mu_1^{p_1}\ra$ at some timer $\mu_1,$ such that
$|rt(\tau_1)-rt(\mu_1)|\le 3d.$ By property (1b),
$rt(\tau_1^{p_1})-rt(\mu_1^{p_1})\le 5d.$ Assume first that
$\latest_{v} = \mu_1^{p_1}.$

At local-time $\tau_1,$ at $q$:
$$\cc_q(\tau_1)=\Cycle-(\tau_1-\tau_1^{p_1})=\Cycle-(rt(\tau_1)-rt(\tau_1^{p_1})).$$
At real-time $t^{\prime\prime},$ $t^{\prime\prime}\ge
rt(\tau_1^{p_1}),$ at $q$:
$$\cc_q(t^{\prime\prime})=\Cycle-(rt(\tau_1)-rt(\tau_1^{p_1}))-(t^{\prime\prime}-rt(\tau_1))=
\Cycle-(t^{\prime\prime}-rt(\tau_1^{p_1})).$$ Similarly at real-time
$t^{\prime\prime},$ $t^{\prime\prime}\ge rt(\mu_1^{p_1}),$ at $v$:
$$\cc_v(t^{\prime\prime})=\Cycle-(t^{\prime\prime}-rt(\mu_1^{p_1})).$$

Thus,$$|\cc_q(t^{\prime\prime})-\cc_v(t^{\prime\prime})|\le 5d.$$

Otherwise, $v$ assigned $\latest_{v}$ as a result of deciding on
some $\la p_2,\_,\mu_2^{p_2}\ra$ at some timer $\mu_2,$
$rt(\mu_2)\le t^{\prime},$ where $\latest_{v}=\mu_2^{p_2}.$ Let
$\tau_2$ be the timer at $q$ when it decided $\la
p_2,\_,\tau_2^{p_2}\ra.$ By the Validity and the
Timeliness-Agreement properties, $|rt(\tau_2)-rt(\mu_2)|\le 3d$ and
$|rt(\tau_2^{p_2})-rt(\mu_2^{p_2})|\le 5d.$

By assumption, $$rt(\tau_1^{p_1})\ge rt(\mu_2^{p_2})\ge
rt(\mu_1^{p_1})\ge rt(\tau_1^{p_1})-5d.$$ At local-time $\tau_1,$ at
$q$:
$$\cc_q(\tau_1)=\Cycle-(\tau_1-\tau_1^{p_1})=\Cycle-(rt(\tau_1)-rt(\tau_1^{p_1})).$$ At local-time $\mu_2,$ at
$v$:
$$\cc_v(\mu_2)=\Cycle-(\mu_2-\mu_2^{p_2}).$$ Let
$t^{\prime\prime}=rt(\tau_1^{\prime\prime})=rt(\mu_1^{\prime\prime}),$
then
$$\cc_q(\tau_1^{\prime\prime})=\Cycle-(t^{\prime\prime}-rt(\tau_1^{p_1})),$$
and
$$\cc_v(\mu_2^{\prime\prime})=\Cycle-(t^{\prime\prime}-rt(\mu_2^{p_2})).$$
Therefore, we conclude
$$|\cc_q(t^{\prime\prime})-\cc_v(t^{\prime\prime})|\le 5d.$$

\end{proof}

\begin{lemma}\label{clm:correct-silencing}
If a correct node $p$ sends a \SP  at some real-time $t_0$ then:
\begin{enumerate}
\item No correct node will invoke \ByzAgreement during the period $[t_0+6d, t_0+\Cycle-d];$
\item No correct node sends a \SP or \PP during that period;
\item The \cc\ counters of all correct nodes expire
within $5d$ of each other at some real-time in the interval
$[t_0+\Cycle-d,t_0+\Cycle+6d].$
\end{enumerate}
\end{lemma}

\begin{proof}
By Lemma~\ref{clm:minus-d} each correct node decides on $p$'s
\SP\!\!. Each correct node that did not update its \latest\
recently, will send a \RE message as a result of this decision.
Since several agreements from different nodes may be executed
concurrently, we need to consider their implication on the resulting
behavior of the correct nodes.

Consider first the case that a correct node reached a decision and
sent \RE before deciding on $p$'s \SP\!\!. If the decision took
place earlier than $t_0-d$ then, by the Timeliness-Agreement
property~(2), it will update it's \latest\ after the decision on
$p$'s \SP\!\!.

By the same Timeliness-Agreement properties, every correct node that
has not sent \RE already, will end up updating its \latest\ and
sending \RE at some time during the interval $[t_0-d,t_0+3d].$ By
$t_0+4d$ no correct node will appear in \proposers\ of any correct
node and until it will send again a \PP message, since its \RE
message will arrive to all non-faulty nodes. Thus, from  time
$t_0+4d$ and until some correct node will send a new \PP message, no
correct node will send \SP message. Moreover, past $t_0+6d$ no
correct node will invoke a \ByzAgreement in Line~D3, because all
correct nodes will not appear also in \recent. Observe that if there
is a \PP message in transit from some correct node $v,$ or if a
correct node $v$ happened to send one just before sending the \RE
message, that \PP will arrive within $d$ of receiving the \RE
message, and therefore by the time that node will be removed from
\recent\ all such messages will arrive and therefore node $v$ will
not be added to \proposers\ as a result of that message later than
$t_0+6d.$

Even though different correct nodes may compute their  $\latest_q$
as a result of different agreements, by the Timeliness-Agreement
properties~(1d) and (2), at time $t_0+6d$ the value of $\latest_q$
satisfies $rt(\latest_q)\in[t_0-d,t_0+6d],$ for every correct node
$q.$

Past time $t_0+6d$ and until $t_0+6d+\BYZdur$ correct nodes may
still decide on values from other agreements that were invoked in
the past by faulty nodes. By the Timeliness-Agreement property~(1c)
no such value result in a \latest\ later than $t_0+6d,$ since no
correct node will invoke \ByzAgreement until some correct node will
send a future \PP message.

Let $t_q$ be the latest real-time a correct node $q$ updated the
calculation of \cc\ because of a \latest\ value in the interval
$[t_0-d,t_0+6d].$ It will send its next \PP message at $t_q+\cc=
t_q+\Cycle-(t_q-rt(\latest_q))=\Cycle+rt(\latest_q)\ge
t_0+\Cycle-d.$ Thus, the earliest real-time a correct node will send
\PP message will be at $t_0+\Cycle-d.$ Until that time no correct
node will send a \PP or \SP message or invoke \ByzAgreement\!\!,
proving (1) and (2).

The bound on \Cycle, implies that during the real-time interval
$[t_0+6d, t_0+\Cycle-\BYZdur]$ there is a window of at least
$14d-6d-d>3d$ with no recording time that refers to it. Denote this
interval by $[t^\prime,t^\prime+B\prime],$ where $B\prime\le
t_0+\Cycle-\BYZdur\le t_0+\Cycle-d.$  The above argument implies
that in the interval $[t^\prime+B\prime,t_0+\Cycle-d]$ no correct
node will update its \latest, and therefore the conditions of
Lemma~\ref{clm:6d} hold.

Thus, the \cc\ counters of all correct nodes expire within $5d$ past
time $t_0+\Cycle-d.$ Looking back at the latest real-time, $t_q\in
[t_0-d,t_0+6d],$ at which a correct node $q$ updated the calculation
of \cc\, the node will send its next \PP message at $t_q+\cc=
t_q+\Cycle-(t_q-rt(\latest_q))=\Cycle+rt(\latest_q)\le
t_0+\Cycle+6d.$ Proving (3).
\end{proof}

Lemma~\ref{clm:decide on-exists} above implies that the nodes will
not deadlock, despite the arbitrary initial states they could have
recovered at. Moreover, by Lemma~\ref{clm:correct-silencing}, once a
correct node succeeds in sending a \SP message, all correct nodes
will converge. We are therefore left with the need to address the
possibility that the faulty nodes will use the divergence of the
initial values of correct nodes to prevent convergence by constantly
causing them to decide and to update their \cc\ counter without
enabling a correct node to reach a point at which it sends a \SP
message.

By Lemma~\ref{clm:decide on-exists} within $\Cycle+4d$ of the time
the system becomes coherent all correct nodes execute Line~E1, thus
within $\Cycle+4d$ from the time the system became coherent. Let
$t_1$ be some real-time in that period by which all non-faulty nodes
executed Line~E1. If any correct node sends a \SP message, then we
are done. Assume otherwise. Since no correct node will invoke
\ByzAgreement for any node more than once within a $\Cycle-11d,$ as
we prove later, there will be at most $f$ decisions between $t_1$
and $t_1+\Cycle-11d.$ Since each decision returns recording times to
nodes that range over at most a $5d$ real-time window, and since
$\Cycle > (10f+16)d,$ there should be a real-time interval $[t_2,
t_2+5d],$ that no recording time refers to any real-time within it.
This reasonings leads to the following lemma.

\begin{lemma}\label{clm:clean-converge}
Assume that no correct node decision results in a recording time
$\tau^p_q$ that refers to real-time $rt(\tau^p_q)$ in the real-time
interval $[t^\prime,t^\prime+5d].$ Then by $t^\prime+\Cycle+4d$ all
correct nodes decide, update their \latest\ and send \RE\!\!, within
$3d$ real-time units of each other.
\end{lemma}

\begin{proof}
By the Timeliness-Agreement property~(1b), any decision that will
take place later than $t^\prime+\BYZdur$ would result in $\latest >
t^\prime.$ By Lemma~\ref{clm:decide on-exists}, by
$t^\prime+\Cycle+4d$ all correct nodes' decisions lead to
$rt(\latest)>t^\prime,$ and by assumption to
$rt(\latest)>t^\prime+5d.$ Let $q$ be the first correct node to
decide and update its \latest\ to a value larger than $t^\prime+5d$
on some $\la p,\_,\tau^p_q\ra$ for $rt(\tau^p_q)>t^\prime+5d,$ at
some real-time $t^{\prime\prime}\ge t^\prime.$ By the
Timeliness-Agreement property~(1d), $t^{\prime\prime}\ge
rt(\tau^p_q).$ Moreover, since the $rt(\tau^p)$ are at most $3d$
apart, by $t^{\prime\prime}+3d$ all correct nodes will decide on
some values and will update the \latest\ value. Therefore, in the
interval $[t^{\prime\prime},t^{\prime\prime}+3d]$ all correct nodes
should update their \latest, with $rt(\latest)\ge t^\prime.$ Thus,
all correct nodes will execute Line~E7 as a result of such
decisions. Therefore, all correct nodes will send a \RE messages
within $3d$ of each other.
\end{proof}

Let $t_2$ be a real time at which the above lemma holds. Let $t_3$
be the real-time past $t_2$ by  which all correct nodes send \RE as
Lemma~\ref{clm:clean-converge} claims. Thus, all correct nodes sent
\RE in the real-time interval $[t_3-3d,t_3]$ and by $t_3+d$ no
correct node will appear in the \proposers\ of any other correct
node.

The final stage of the proof is implied from the following lemma.

\begin{lemma}\label{clm:all-reset}
If all correct nodes send a \RE  in the period $[t_0,t_0+3d]$ then:
\begin{enumerate}
\item No correct node will invoke \ByzAgreement during the period $[t_0+6d, t_0+\Cycle-\BYZdur];$
\item No correct node sends a \SP or \PP during that period;
\item The \cc\ counters of all correct nodes expire
within $5d$ of each other at some real-time in the interval
$[t_0+\Cycle-\BYZdur,t_0+\Cycle+6d].$
\end{enumerate}
\end{lemma}

\begin{proof}
By real-time $t_0+4d$ all correct nodes will receive all the $n-f$
\RE messages and will remove the correct nodes from \proposers. Past
that time and until some correct node will send a \PP in the future,
no correct node will send a \SP message. Similarly, past $t_0+6d$
and until some correct node will send a \PP in the future no correct
node will invoke a \ByzAgreement in Line~D3.

At that time the range of \cc\ may be in $[\Cycle-\BYZdur,\Cycle],$
since, by the Timeliness-Agreement property~(1d), faulty nodes may
bring the correct nodes to decide on values that are at most
$\BYZdur$ in the past.

Until $t_0+6d+\BYZdur,$ correct nodes may still decide on values
from other agreements invoked by faulty nodes. By the
Timeliness-Agreement property~(1c), until some correct node will
invoke a $\ByzAgreement,$ no correct node will happen to decide on
any message with $rt(\tau^\prime)\ge t_0+6d$ (latest possible
recording time).

Let $t_q$ be the latest real-time a correct node $q$ updated the
calculation of \cc\ at some time during the interval $[t_0,
t_0+6d].$ It will send its next \PP message at $t_q+\cc=
t_q+\Cycle-(t_q-rt(\latest_q))=\Cycle+rt(\latest_q).$ By
Timeliness-Agreement property~(1d), and because the computation of
$\latest_q$ takes place in the interval $[t_0,t_0+6d]$ we conclude
that interval $rt(\latest_q)\ge t_0+ \Cycle-\BYZdur.$ Thus, $t_q+\cc
\ge t_0+ \Cycle-\BYZdur.$ Thus, the earliest time a correct node
will send a \PP message will be at $t_0+\Cycle-\BYZdur.$ Until that
time no correct node will send a \PP or \SP message or invoke
\ByzAgreement\!\!, proving (1) and (2).

The bound on \Cycle, implies that during the real-time interval
$[t_0+6d, t_0+\Cycle-\BYZdur]$ there is a window of at least
$14d-6d-d>3d$ with no recording time that refers to it. Denote this
interval by $[t^\prime,t^\prime+B^\prime],$ where $B\prime\le
t_0+\Cycle-\BYZdur\le t_0+\Cycle-d.$  The above argument implies
that in the interval $[t^\prime+B^\prime,t_0+\Cycle-d]$ no correct
node will update its \latest, and therefore the conditions of
Lemma~\ref{clm:6d} hold.

Thus, the \cc\ counters of all correct nodes expire within $5d$ past
time $t_0+\Cycle-d.$ Looking back at the latest real-time, $t_q,$ at
which a correct node $q$ updated the calculation of \cc, since it
took place in the interval $[t_0-d,t_0+6d]$ and that $rt(\latest_q)$
cannot be larger than the time at which it is computed, the node
will send its next \PP message at $t_q+\cc=
t_q+\Cycle-(t_q-rt(\latest_q))=\Cycle+rt(\latest_q)\le
t_0+\Cycle+6d.$ Proving (3).

\end{proof}

\begin{corollary}\label{clr:all-reset}
In the conditions of Lemma~\ref{clm:all-reset}, if no correct node
invoked \ByzAgreement in the interval $[t_0-\BYZdur,t_0-B]$ then the
bound of $t_0+\Cycle-\BYZdur$ in Lemma~\ref{clm:all-reset} can be
replaced by $t_0+\Cycle-B-2d.$
\end{corollary}

\begin{proof}
By the Timeliness-Agreement property~(1c) no decision can return a
recording time that is earlier by more than $2d$ from an invocation
of \ByzAgreement by a correct node. Therefore, in the proof of
Lemma~\ref{clm:all-reset} the minimal value for $\latest_q$ for any
correct node $q$ can be $t_0-B-2d.$ Let $t_q$ be the latest
real-time a correct node $q$ updated the calculation of \cc\ in the
interval $[t_0, t_0+6d].$ It will send its next \PP message at
$t_q+\cc= t_q+\Cycle-(t_q-rt(\latest_q))=\Cycle+rt(\latest_q)\ge
\Cycle-B-2d.$ Thus, the earliest real-time a correct node will send
\PP message will be at $t_0+\Cycle-B-2d.$ Until that time no correct
node will send a \PP or \SP message or invoke \ByzAgreement\!\!.
Thus the bound of $t_0+\Cycle-\BYZdur$ in Lemma~\ref{clm:all-reset}
can be replaced by $t_0+\Cycle-B-2d.$
\end{proof}

We can now state the ``fixed-point'' lemma:

\begin{lemma}\label{clm:all-pp}
If the \cc\ counters of all correct nodes  expire in the period
$[t_0,t_0+5d]$ and no correct node sent \SP in
$[t_0-(\Cycle-8d),t_0]$ and no correct node invoked \ByzAgreement in
$[t_0-(\BYZdur+6d),t_0]$ then:
\begin{enumerate}
\item All correct nodes invoke a pulse within $3d$ real-time units of each other before $t_0+9d$;
\item There exists a real-time $\bar t_0,$ $t_0+\Cycle-2d\le \bar t_0\le
t_0+\Cycle+12d$ for which the conditions of Lemma~\ref{clm:all-pp}
hold by replacing $t_0$ with $\bar t_0.$
\end{enumerate}
\end{lemma}

\begin{proof}
Assume first that a correct node decided in $[t_0,t_0+6d].$ Let $q$
be the first such correct node to do so, at some real-time $t_q.$ By
the Timeliness-Agreement property~(1c), and since no correct node
has invoked \ByzAgreement in $[t_0-(\BYZdur+6d),t_0],$ the recording
time needs to be in the interval $[t_0-2d,t_q].$  By
Timeliness-Agreement property~(1a), in the interval $[t_q,t_q+3d]$
all correct nodes will decide, and the decision of all correct nodes
will imply updating of \latest\ and the conditions for invoking a
pulse hold.

Moreover, in the interval $[t_q,t_q+3d]$ the preconditions of
Lemma~\ref{clm:all-reset} holds. Using Corollary~\ref{clr:all-reset}
for $B=0$ we obtain the bounds of no \SP in $[t_q+6d,
t_q+\Cycle-2d],$ an interval of $\Cycle-8d,$ and all \cc\ expire
within $5d$ in the interval $[t_q+\Cycle-2d,t_q+\Cycle+6d].$ Since
$t_q\in [t_0,t_0+6d],$ we conclude that for $\bar
t_0\in[t_q+\Cycle-2d,t_q+\Cycle+12d]$ the conditions of the lemma
hold.

Otherwise, no correct node decided in $[t_0,t_0+6d].$ This implies
that all correct nodes will end up sending their \PP by $t_0+5d$ and
a correct node will send \SP by $t_0+6d.$
Lemma~\ref{clm:correct-silencing} completes the proof in a similar
way.
\end{proof}

Observe that once Lemma~\ref{clm:all-pp} holds, it will hold as long
as the system is coherent, since its preconditions continuously
hold. So to complete the proof of the theorem we need to show that
once the system becomes coherent, the preconditions of
Lemma~\ref{clm:all-pp} will eventually hold.

Denote by $\tilde t$ the real-time at which the system became
coherent. By Lemma~\ref{clm:decide on-exists} by $\tilde
t+\Cycle+4d$ all correct node executes Line~E1. Let $t_1$ be some
real-time in that period by which all correct nodes executed
Line~E1. If any correct node sends a \SP message, then by
Lemma~\ref{clm:correct-silencing} the precondition to
Lemma~\ref{clm:all-pp} hold.

Assume otherwise. By the Timeliness-separation property there are no
concurrent agreements associated with the same sender of \SP
message. Since the separation between decisions is at least $5d,$
every correct node will be aware of a decision before invoking the
next \ByzAgreement and therefore, the test in Line~D2 will eliminate
having more than a single decision per sending node within
$\Cycle-11d.$ Since $\Cycle
> (10f+16)d,$ there will be at most $f$ decisions between $t_1$ and
$t_1+\Cycle-11d.$ Since each decision returns recording times to
nodes that range over at most $5d$ real-time window, there should be
a real-time interval $[t_2, t_2+5d],$ that no recording time of any
correct node refers to any real-time within it. Note that $t_2\le
t_1+\Cycle-11d-5d\le \tilde t+2\cdot\Cycle-12d.$ By
Lemma~\ref{clm:clean-converge}, by $\tilde
t+2\cdot\Cycle-12d+\Cycle+4d\le \tilde t+3\cdot\Cycle-8d$ there
exist a $t_3$ such that all correct nodes sends \RE in the interval
$[t_3-3d,t_3].$ Thus, the preconditions to Lemma~\ref{clm:all-reset}
hold. Thus, by $\tilde t+3\cdot\Cycle-8d-3d+\Cycle+6d=\tilde
t+4\cdot\Cycle-5d$ the preconditions to Lemma~\ref{clm:all-pp} hold
because either a correct node has sent \SP before that or from
Lemma~\ref{clm:all-reset}.

Thus the system converges within less than $4\cdot\Cycle$ from a
coherent state. One can save one \Cycle\ in the bound by overlapping
the first one with the second one when the non-faulty nodes are not
being considered correct.

From that time on, all correct nodes will invoke pulses within $3d$
of each other and their next pulse will be in the range stated by
Lemma~\ref{clm:all-pp}. The Lemma immediately implies that the bound
on $\cyclemax$ is $\Cycle+9d.$ Similarly, it claims that past
$t_0+9d$ no \PP\ will be sent before $t_0+\Cycle-2d,$ thus
potentially the shortest time span between pulses at a node is
$\Cycle-11d.$ This implies that $\cyclemin=\Cycle-11d.$ Moreover,
the discussion also implies that:

\begin{lemma}\label{clm:1-pulse}
Once the conditions of Lemma~\ref{clm:all-pp} hold, no correct node
will invoke more than a single pulse in every $\cyclemin$ real-time
interval. It will invoke at least one pulse in every $\cyclemax$
real-time interval.
\end{lemma}

This concludes the Convergence requirement with $\sigma = 3d,$
 since the correct nodes will always invoke pulses
within $3d$ real-time units of each other. This completes the proof
of Theorem~\ref{thm:converge}.}
\end{proof}

\begin{theorem}\label{thm:closure}
(\emph{Closure}) If the system is in a synchronized pulse\_state at
time $t_s,$ then the system is in a synchronized pulse\_state at
time $t,$ $t \ge t_s.$
\end{theorem}

\begin{proof}
Let the system be in a synchronized pulse\_state at the time
immediately following the time the last correct node sent its \PP
message. Thus, all correct nodes have sent their \PP messages. As a
result, all will invoke their pulses within $3d$ of each other, and
will reset $\cc$ to be at least $\Cycle-2d.$  The faulty nodes may
not influence the \cycle\ length to be shorter than \cyclemin\ or
longer than \cyclemax.
\end{proof}

Thus we have proved the main theorem:

\begin{theorem}\label{thm:pulse-ss}
\emph{(Convergence and Closure)} The $\Pulsealg$ algorithm solves
the Self-stabilizing Pulse Synchronization Problem if the system
remains coherent for at least $4$ cycles.
\end{theorem}

\begin{proof} Convergence follows from Theorem~\ref{thm:converge}. The
first Closure condition follows from Theorem~\ref{thm:closure}. The
second Closure condition follows from Lemma~\ref{clm:1-pulse}.
\end{proof}

Since we defined non-faulty to be considered correct within 2
cycles, we conclude:

\begin{corollary}
From an arbitrary state, once the network become correct and $n-f$
nodes are non-faulty, the $\Pulsealg$ algorithm solves the
Self-stabilizing Pulse Synchronization Problem if the system remains
so for at least $6$ cycles.
\end{corollary}

\begin{lemma}\label{clm:to-be-correct} (\emph{Join of recovering nodes}) If the system is in synchronized state,
a recovered node becomes synchronized with  all correct nodes within
$\Delta_{node}$ time.
\end{lemma}

\begin{proof}
The proof follows the arguments used in the proofs leading to
Theorem~\ref{thm:pulse-ss}.  Within a \cycle\ of non-faulty behavior
of the recovering node it clears its variable and data structures of
old values. Within \cyclemax\ it will synchronize with all other
correct nodes, though it might not issue a pulse if it issued one in
the first \Cycle. But by the end of $\Delta_{node}$ its \cc\ will
synchronize with all the correct nodes and will consequently produce
the next pulse in synchrony with them.
\end{proof}


\end{document}